\documentclass{ws-p8-50x6-00}

\usepackage{psfrag}

\begin{document}

\title{Twist--3 effects in Deeply Virtual Compton Scattering made simple}

\author{C.~WEISS}

\address{Institut f\"ur Theoretische Physik \\
Universit\"at Regensburg, D--93053 Regensburg, Germany \\
E-mail: christian.weiss@physik.uni-regensburg.de}  

\maketitle

\abstracts{We show that electromagnetic gauge invariance requires 
a ``spin rotation'' of the quarks in the usual twist--2 
contribution to the amplitude for Deeply Virtual Compton Scattering. This
rotation is equivalent to the inclusion of certain kinematical twist--3
(``Wandzura--Wilczek type'') terms, which have been derived 
previously using other methods. The new representation of the twist--3 
terms is very compact and allows for a simple physical interpretation.}

Deeply Virtual Compton Scattering (DVCS), $\gamma^*(q) + N(p) 
\rightarrow \gamma (q') + N(p')$ at large $q^2$ and finite $t = (p' - p)^2$, 
is the simplest process which could probe the generalized parton 
distributions (GPD's) in the nucleon. New experimental results for spin 
and charge asymmetries of the cross section have been reported at this 
meeting, allowing for a first comparison of GPD models 
with data.\cite{hermes,clas}

The crucial property of DVCS (and a number of other hard electroproduction
processes) is that the amplitude can be factorized in a hard photon--quark 
amplitude, and a soft matrix element containing the relevant information 
about the structure of the nucleon. Technically, this factorization 
can be accomplished using QCD expansion techniques familiar from the 
theory of deep--inelastic scattering. Originally only the 
contribution from twist--2 operators was included.\cite{dvcs} 
It was realized that in this approximation the amplitude is not
transverse (electromagnetically gauge invariant); the violation
is proportional to the transverse component of the momentum 
transfer, which is not suppressed at large $q^2$. 
A gauge invariant amplitude up to terms $O(t/q^2)$ is obtained 
by including certain ``kinematical'' twist--3 contributions. These have 
been derived in various approaches: Momentum--space collinear 
expansion,\cite{Anikin:2000em} coordinate--space light cone 
expansion,\cite{Belitsky:2000vx,Radyushkin:2000jy}
and a parton--model based approach.\cite{Penttinen:2000dg}
In the usual formulation the twist--3 terms are
parametrized by auxiliary GPD's given by certain integrals over the
basic twist--2 GPD's, much like the Wandzura--Wilczek part of
the spin structure function $g_2 (x)$ in inclusive DIS. In addition
to restoring gauge invariance of the twist--2 contribution, the
twist--3 terms give rise to new helicity amplitudes and strongly 
influence the predictions for the spin and charge asymmetries of 
the DVCS cross section.\cite{observables}

In this talk I would like to point out that the kinematical
twist--3 terms in the DVCS amplitude have a simple physical 
interpretation as being due to a ``spin rotation'' applied to the
twist--2 quark density matrix in the nucleon.\cite{Radyushkin:2001ap}
This allows for a very compact representation of the twist--3 effects.
Most important, it shows that, in spite of the apparent complexity
of the amplitude at twist--3 level, DVCS is still a ``simple'' 
process. The results reported here have been obtained in collaboration 
with A.~V.~Radyushkin (Jefferson Lab and Old Dominion U.)

%
%
\begin{figure}[t]
\begin{tabular}{ccc}
\includegraphics[width=4cm,height=3.7cm]{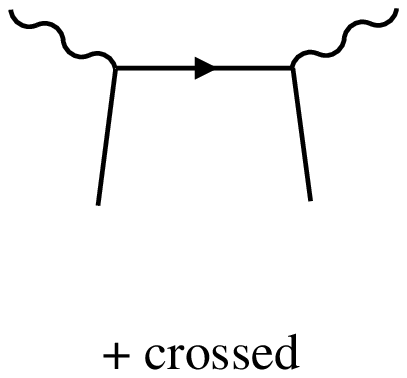}
& \hspace{.5cm} &
\includegraphics[width=4.4cm,height=3.7cm]{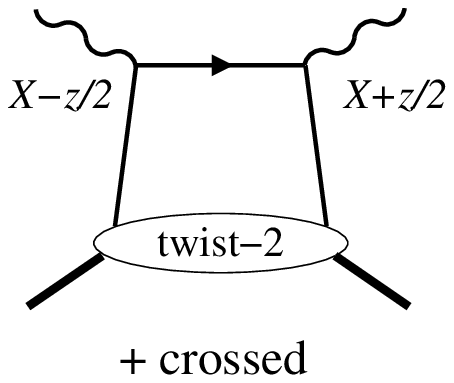}
\\[.2cm]
{\large (a)} & & {\large (b)}
\end{tabular}
\caption[]{}
\label{fig_1}
\end{figure}
Consider virtual Compton scattering off an electron in QED at 
tree level, see Fig.~\ref{fig_1}a. It is well--known that transversality 
of the amplitude, $q_\mu T_{\mu\nu} = 0$ and $T_{\mu\nu} q'_\nu = 0$,
requires not only the Ward identities relating the electromagnetic
vertex and the free--field propagator, but also the on--shell conditions 
for the external particles, {\it i.e.}, the Dirac equations 
for the electron spinors.

Turning now to DVCS off a hadron, the twist--2 contribution to the 
amplitude in QCD is given by exactly the same diagrams as Fig.~\ref{fig_1}a,
describing virtual Compton scattering off a free quark, only the 
wave functions of the initial and final particle have been replaced by 
the transition matrix element of the appropriate non-local
quark/antiquark density matrix between the hadronic states, 
see Fig.~\ref{fig_1}b. The twist--2 part of the latter is defined as
\begin{equation}
M_{ij} (z|X)^{\mbox{\scriptsize twist--2}}
\;\; = \;\; \int_0^1 d\lambda \left( \gamma_\sigma \right)_{ij}
\frac{\partial}{\partial z_\sigma}
\langle p' | \; \bar\psi (X-\lambda z/2) \hat z \psi (X+\lambda z/2) \; 
| p \rangle
\label{M_twist_2}
\end{equation}
plus a similar contribution with 
$\gamma_\sigma \rightarrow \gamma_5 \gamma_\sigma$ and
$\hat z \rightarrow \hat z\gamma_5$. The density matrix is presented 
here in perhaps somewhat unusual form, in coordinate space, with the 
quark/antiquark ``ends'' located at $X \pm z/2$ ($X$ is the center 
coordinate, $z$ the separation);
$i$ and $j$ are the Dirac spinor indices. Here $\bar\psi$ and $\psi$ are 
the quark fields (we omit the flavor labels), and the bilinear operator 
is really a traceless QCD string operator, see Ref.\cite{Radyushkin:2001ap} 
for details. What is important is that this twist--2 density matrix 
{\em does not satisfy} the free--field Dirac equations with respect to the 
quark/antiquark ``ends''; the violation is proportional to the momentum 
transfer $\Delta = p' - p$. The reason is, simply put, that in the 
twist--2 operator in Eq.(\ref{M_twist_2}) the quark 
spin is projected on a {\it fixed} direction, determined by the vector $z$, 
while the Dirac equations require that the spin projection {\it changes} 
between the two ends in accordance with the momentum transfer between the
quark lines.\footnote{In the usual collinear expansion around a fixed 
light--like direction, the vector operator in Eq.(\ref{M_twist_2}) would 
have a large ``plus'' component, while the quark/antiquark ends have 
transverse momenta because of $\Delta_\perp \neq 0$.} As a consequence, the 
twist--2 part of the DVCS amplitude alone is not electromagnetically gauge 
invariant; the amplitude violates transversality by terms proportional
to $\Delta$.

%
%
\begin{figure}[t]
\hspace{1cm} \includegraphics[width=7.5cm,height=3.7cm]{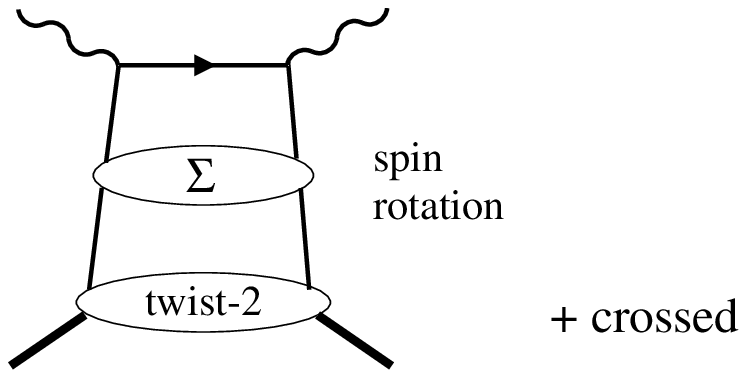}
\caption[]{}
\label{fig_spinrot}
\end{figure}
It is not difficult to see what must be done in order to fix this problem.
We must rotate the spin projection of the quarks in the density matrix
(\ref{M_twist_2}) such as to align it with the momenta of the
incoming and outgoing quark ends. This is achieved by a 
position--dependent rotation with a matrix\cite{Radyushkin:2001ap}
\begin{equation}
\Sigma\,(z/2) \;\; \equiv \;\; \exp \left[ -\frac{i}{4} 
z_\alpha \sigma_{\alpha\beta} \Delta_\beta \right] .
\label{Sigma}
\end{equation}
The modified density matrix is ($\bar \lambda \equiv 1 - \lambda$)
\begin{eqnarray}
M_{ij} (z|X)^{\mbox{\scriptsize rot}}
&=& \int_0^1 d\lambda \left[ \Sigma\,(\bar \lambda z/2 )
\; \gamma_\sigma \;
\Sigma\,(\bar \lambda z/2 ) \right]_{ij}
\nonumber \\
&& \times \frac{\partial}{\partial z_\sigma}
\langle p' | \; \bar\psi (X-\lambda z/2) \hat z \psi (X+\lambda z/2) \; 
| p \rangle
\label{M_rot}
\end{eqnarray}
plus the same with $\gamma_\sigma \rightarrow \gamma_5 \gamma_\sigma$
and $\hat z \rightarrow \hat z\gamma_5$. This ``rotated'' form 
satisfies the Dirac equations with respect to the external ends,
up to terms proportional to $t$, see Ref.\cite{Radyushkin:2001ap} 
for details. As a result, the DVCS amplitude
obtained with Eq.(\ref{M_rot}) is gauge invariant 
up to terms of order $O(t/q^2)$. Schematically,
our modification of the twist--2 contribution to the DVCS amplitude
can be represented as in Fig.\ref{fig_spinrot}, with the spin rotation
as an ``intermediate step'' between the twist--2 density matrix 
and the free quark Compton amplitude.

In the terminology of the light cone expansion, the spin rotation
of Eq.(\ref{M_rot}) amounts to the inclusion of
certain twist--3 operators, which, however, are completely given
in terms of total derivatives of twist--2 operators
(``kinematical twist--3''). When substituting parametrizations
for the basic twist--2 matrix elements, Eq.(\ref{M_rot}) 
reproduces the Wandzura--Wilczek type relations
for the  twist--3 GPD's, which were derived previously using
other techniques.\cite{Radyushkin:2000jy} Thus, all the
complexity of the kinematical twist--3 effects in DVCS can
be reduced to the simple spin rotation of Eq.(\ref{M_rot}).

The effect of kinematical twist--3 terms on DVCS observables have
been discussed in the literature.\cite{observables} 
The twist--3 terms affect in particular the spin and charge asymmetries 
of the cross section. The spin rotation representation could be helpful 
in developing a more intuitive understanding of the twist--3 
effects in DVCS observables. This problem certainly deserves further 
study.

C.W.\ is supported by a Heisenberg Fellowship from 
Deutsche Forschungsgemeinschaft (DFG).

\end{document}